\documentclass[12pt]{article}
\usepackage{graphics}
\usepackage{epsfig}
\usepackage{subfigure}
\usepackage{graphics}
\usepackage{amsmath}
\usepackage{amsfonts}
\usepackage{amssymb}
\usepackage{url}
\usepackage{hyperref}
\usepackage{color}

\newcommand{\be}{\begin{equation}}
\newcommand{\ee}{\end{equation}}
\newcommand{\bea}{\begin{eqnarray}}
\newcommand{\eea}{\end{eqnarray}}
\newcommand{\bref}[1]{(\ref{#1})}
\begin{document}
\begin{titlepage}
\begin{flushright}
\today
\end{flushright}
\vspace{4\baselineskip}
\begin{center}
{\Large\bf 
Blandford-Znajek Effect and Neutrino Pair Annihilation as the Central Engines of Gamma Ray Bursts 
 and Blazars}
\end{center}
\vspace{1cm}
\begin{center}
{\large
Takeshi Fukuyama
\footnote{\tt E-mail:fukuyama@rcnp.osaka-u.ac.jp},
}
\end{center}
\vspace{0.2cm}
\begin{center}
{\small \it ${}^a$Research Center for Nuclear Physics (RCNP),
Osaka University, \\Ibaraki, Osaka, 567-0047, Japan}
\end{center}
\vskip 5mm

\begin{abstract}
The central engine of gamma-ray burst is considered.  Blandford-Znajek (BZ) process is reconsidered under the asymtotic magnetic condition ${\bf B}=B_0\hat{z}$ (rotation axis) at $r\gg r_H$. Another scenario of $\nu \overline{\nu} \to e^+ e^-$ driven jet are combined to BZ process to compliment it.
\end{abstract}
\end{titlepage}
\section{Introduction}
Gamma-Ray Bursts (GRBs) are short and intense pulses of $\gamma$-rays.
The bursts last from a fraction of a second to several hundred seconds. See Refs.\cite{GRB1,GRB2,GRB3, Nagataki} for reviews. In this letter, we consider the central engine of GRBs whose typical duration times range from several seconds to several hundred seconds (so called long GRBs). These mechanism may be applied to the blazars \cite{Agudo}.
There are two categories of the central engine's mechan isms of GRB jets, magnetically driven jets due to the Blandford-Znajek (BZ) effect \cite{Blandford} and the neutrino driven jets due to $\nu\overline{\nu}\to e^+e^-$ reactions \cite{Woosley, Popham, Asano1, Asano2}, which are based on a special neutron star (magnetar) and black holes (BHs).
However, both scenarios seem to need to incorporate the other elements for their completeness.
That is, magnetically driven jet scenario needs charged sources to support its electric current, and to clarify the detailed distribution of the magnetic field around the source. For the $\nu\overline{\nu}$-driven mechanism, we needs magnetic field to convert the produced $e^+e^-$ energy to finally $\gamma$ rays.  The purpose of the present letter is to clarify these points. 

 As is well known, the most energy of gravitational potential by collapsing is emitted via $\nu\overline{\nu}$ pairs.
We consider the central engine of GRB via BZ process supplemented with the charge source of $e^-e^+$ due to pair annihilation of $\nu\overline{\nu}$ coming from optically thin accretion disc along the rotation axis. 

In this letter we adopt $\hbar=c=1$ units unless otherwise specified.

\section{BZ Effect}
In this section, we review the BZ effect and rederive its power taking the detailed distribution of magnetic field into consideration.
We consider the Kerr black hole (BH) whose mass is $M$,
\bea
ds^2&=&-\frac{\Delta-a^2\sin^2\theta}{\Sigma}dt^2-\frac{2r_gra}{\Sigma}\sin^2\theta dtd\varphi\nonumber \\
&+&\frac{A}{\Sigma}\sin^2\theta d\varphi^2+\frac{\Sigma}{\Delta}dr^2+\Sigma d\theta^2,
\eea
where 
\bea
r_g&=&2GM, ~~a=\frac{J}{M} \\
A&=&(r^2+a^2)^2-\Delta a^2\sin^2\theta \\
\Delta&=& r^2-r_gr+a^2,~~\Sigma=r^2+a^2\cos^2\theta.
\eea

This metric is stationary and axially symmetric. That is, it has two Killing vectors
\be
k_\mu=\delta_\mu^t, ~~l_\mu=\delta_\mu^\varphi
\ee
The sum of energy-momentum tensor of matter $T_{\mu\nu}^{(m)}$ and that of field $T_{\mu\nu}^{(f)}$ is conserved
\be
\left(T^{(m)\mu\nu}+T^{(f)\mu\nu}\right)_{;\nu}=0,
\ee
where $;$ indicates covariant derivative. $T_{\mu\nu}^{(f)}$ satisfies the Maxwell equation
\be
{T^{(f)\mu\nu}}_{;\nu}=-F^{\mu\nu}j_\nu.
\ee
In this paper we assume 
\be
|T^{(f)\mu\nu}| \gg |T^{(m)\mu\nu}|
\label{TF}
\ee
and, therefore, we obtain the force free condition
\be
F^{\mu\nu}j_\nu=0.
\label{FF}
\ee
In this letter, we do not adopt the magneto-hydro dynamic (MHD) condition, that is,
\be
{\bf E}+{\bf v}\times{\bf B}\neq 0.
\label{MHD}
\ee
and ${\bf E}\cdot{\bf B}\neq 0$ (See Eq.\bref{EdotB}).
Namely we do not assume highly conductive supernovae remnant and may expect to apply our scenario to more general objects like blazars.

Here we consider the homogeneous magnetic field ${\bf B}=B_0\hat{z}$ at $r\gg r_H$ in the zero angular momentum observer (ZAMO) frame \cite{Wald, King}, where $r_H$ is the BH radius
\be
r_H=\frac{r_g}{2}+\sqrt{\left(\frac{r_g}{2}\right)^2-a^2}.
\ee
This distribution is free from the singularity at $r_H$ unlike the stationary axisymmetric electromagnetic field \cite{Petterson, Prasanna}. 
The poloidal components of magnetic field
can be written as
\bea
B_r=A_{\varphi,\theta}&=&\frac{B_0}{A^{1/2}\Sigma^2}\left\{(r^2+a^2)\left[(r^2-a^2)(r^2-a^2\cos^2\theta)+2a^2r(r-\frac{r_g}{2})(1+\cos^2\theta)\right]\right. \nonumber\\
&&\left. -a^2\Delta\Sigma\sin^2\theta\right\}\cos\theta,
\label{Br}
\eea
and
\bea
B_\theta&=&\frac{-N\Delta^{1/2}}{A^{1/2}\Sigma^2}\left\{a^2\left[2r(r^2-a^2)\cos^2\theta-(r-M)(r^2-a^2\cos^2\theta)(1+\cos^2\theta)\right]\right.\nonumber\\
&&\left.+(r^2+a^2)\Sigma r\right\}\sin\theta.
\eea
Then, $B_\theta\approx 0$ around $r\approx r_H$.
Thus the magnetic fields decreases as the angular momentum of BH increases.
The BZ power is given by \cite{Blandford, Thorne}
\be
\Delta P_{BZ}=\frac{\Omega_F(\Omega_H-\Omega_F)}{4\pi}\frac{A}{\Sigma}\sin^2\theta B_r\Delta\Psi.
\ee
Here $\Omega_F$ and $\Omega_H$ are the angular velocities of field and BH, respectively, and $\Delta\Psi$ is the magnetic flux over the annual surface between $\theta$ and $\theta+\Delta\theta$,
\be
\Delta \Psi=B_r\Delta A_{ann}=4\pi r^2 \sin\theta\Delta\theta B_r.
\ee
On the other hand, $\Omega_H$ defined by
\be
\Omega_H=\frac{a}{2GMr_H}.
\ee
$P_{BZ}$ is maximized as $\Omega_F=\frac{\Omega_H}{2}$, adopting this value hereafter.
As is shown from Eq.\bref{Br}, $a$ increases, $B_r^2$ decreases and $\Omega_H^2$ increases. The total power $P_{BZ}$
is obtained by integrating $\theta$ from $0$ to $\pi/2$ and represented as Fig.1.
\vspace{0.5cm}
\begin{figure}[h]
\begin{center}
\includegraphics[scale=0.7]{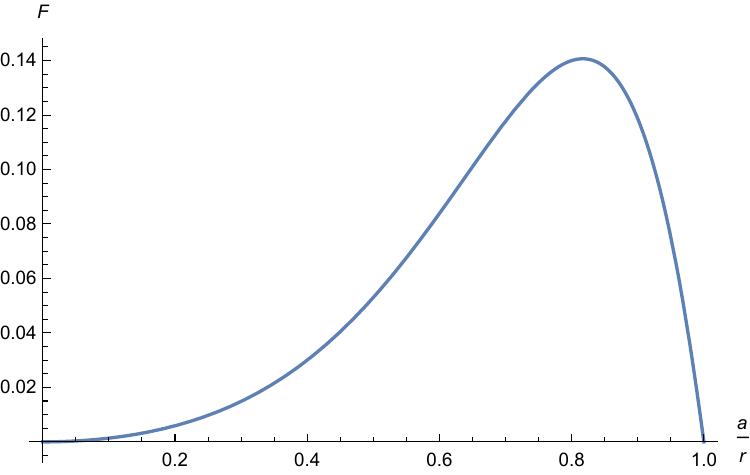}
\caption{The total power of BZ effect: The horizontal axis is $a/r$, and the vertical axis is $F(a/r)$ normalized as Eq.\bref{Fb}.
  }
\label{fig:power}
\end{center}
\end{figure}

Finally we obtain
\be
P_{BZ}=\frac{1}{32}(GM)^2B_0^2F(a/r)\approx 2\times \left(\frac{M}{M_\odot}\right)^2\left(\frac{B}{10^{15}~\mbox{Gauss}}\right)^2F(a/r)\times 10^{49}~\mbox{erg/s}.
\label{Fb}
\ee
As is shown in Fig.1, $F(a/r)$ is maximally $0.14$ at $a/r\approx 0.8$.
This $P_{BZ}$ ia also described as
\be
P_{BZ}\approx 2\times \left(\frac{M}{10^8M_\odot}\right)^2\left(\frac{B}{10^{4}~\mbox{Gauss}}\right)^2F(a/r)\times 10^{44}~\mbox{erg/s}.
\ee
This may be referred as the case of blazars \cite{Agudo}.

The well known result \cite{ Blandford3, HKLee}
\be
P_{BZ}\approx 1.7\times 10^{50} \left(\frac{a}{GM}\right)^2\left(\frac{M}{M_\odot}\right)^2\left(\frac{<B>}{10^{15}\mbox{Gauss}}\right)^2 \mbox{erg/s}
\ee
is larger than our result by roughly the magnitude $10^2$. Here $B(\theta)$ is approximated as constant $<B>$. As we have explained in Eq.\bref{Fb}, $P_{BZ}$ depends on $M$ and $B$ only and not on $a$ independently.
Also we have not assumed ideal MHD condition Eq.\bref{MHD} but obtain \cite{King}
\bea
{\bf E}\cdot{\bf B}&=&\frac{B^2a\cos\theta}{\Sigma^4}\left\{\Delta r\Sigma^2\sin^2\theta + [(r^2-a^2)(r^2-a^2\cos^2\theta)+2a^2r(r-GM)(1+\cos^2\theta)]\right. \nonumber\\
&&\times\left.[2r(r^2-a^2)\cos^2\theta-(r-GM)(r^2-a^2\cos^2\theta)(1+\cos^2\theta)]\right\}.
\label{EdotB}
\eea
The non-vanishing ${\bf E}\cdot{\bf B} >0 (<0)$ indicates that negative (positive) charged particles are pulled into BH with $B_0>0$.
The signature on $r_H$ is determined by that of the second $[..]$ of Eq.\bref{EdotB} and by
\be
a^2x^2+2r_H(r_H+GM)x-r_H^2=0
\ee
with $x=\cos^2\theta.$
Thus ${\bf E}\cdot{\bf B}>0$ between $0^\circ$ and $40^\circ$ for $a\approx 0~(60^\circ$ for $a=GM$), where positive charges are accelerated outward for $B_0>0$. These charged particles may be supplied by the
$e^+e^{-}$ driven by $\nu\overline{\nu}$ pairs, which will be given in the next section.
This is the alternative scenario to the polar gap model \cite{Goldreich, Ruderman, Beskin, Hirotani}.

\section{$\nu\overline{\nu}\to e^+e^-$ Reactions}
In order to maintain the current flows we need charged particles outside the horizon \cite{HKLee}. As the source of this current and complementary (or even alternative) power to the BZ process, we consider $e^+e^-$ production due to $\nu\overline{\nu}$ which come from the accretion disk \cite{Popham, Asano1, Asano2}.

\be
\nu +\overline{\nu}~\to~e^++e^-~\to 2\gamma.
\ee
 The detailed calculations are given in \cite{Asano1, Asano2} and here we give the result only. 
The energy deposition rate (EDR) due to $\nu\overline{\nu}$ annihilation along the rotation axis within $\theta<10^\circ$ is given by Eq.(11) of \cite{Asano2}
\be
\frac{dE(r)}{dtdV}=\frac{21\pi^4}{4}\zeta(5)\frac{KG_F^2}{h^6c^5}(kT_{eff}(3r_g))^9F(r)
\label{EDR}
\ee
with
\bea
F(r)&=&\frac{1}{T^9_{eff}(3r_g)}\left(\frac{r^2+a^2}{\Delta}\right)^4\int_{\theta_m}^{\theta_M}d\theta_\nu\sin\theta_\nu\times\int_{\theta_m}^{\theta_M}d\theta_{\overline{\nu}}\sin\theta_{\overline{\nu}}\int_0^{2\pi}d\varphi _\nu\int_0^{2\pi}d\varphi_{\overline{\nu}}T_0^5(R_\nu)\nonumber\\
&\times&T_0^4(R_{\overline{\nu}})\left[1-\sin\theta_\nu\sin\theta_{\overline{\nu}}\cos(\varphi_\nu-\varphi_{\overline{\nu}})-\cos\theta_\nu\cos\theta_{\overline{\nu}}\right]^2.
\label{F}
\eea
Here we have written $h=2\pi\hbar$ and $c$ explicitly for clarity. The definitions of $\theta_m,~\theta_M$ are due to \cite{Asano2}.
Integration of Eq.\bref{EDR} over the volume $dV=(r^2+a^2\cos^2\theta)\sin\theta drd\theta d\varphi$ gives the energy deposition per unit world time for a distant observer. 
\be
\frac{dE}{dt}\approx 4.41 \times 10^{48}\left(\frac{d\theta}{10^\circ}\right)^2\left(\frac{kT_{eff}(3r_g)}{10\text{MeV}}\right)^9\times\left(\frac{r_g}{10\text{Km}}\right)^3\int_{1.5}^{10}G(\hat{r})d\hat{r}~\text{erg/s},
\label{EDR1}
\ee
where $G(r)=F(r)(r^2+a^2)/r_g^2$. 
In table 1 of \cite{Asano2} we list the integral of $G(r)$ over $r=1.5-10r_g$. We have not considered the EDR $r<1.5 r_g$  since in this region the baryon contamination occurs severely and reabsorption rate of the deposition to BH is large. Also for $r>10r_g, ~F(r)$ essentially vanishes. Also from the baryon contamination we have restricted EDR within $10^\circ$ along the rotation axis. Thus EDR becomes
\begin{equation}
\frac{dE}{dt} \simeq 2.23 \times 10^{51}
\left( \frac{d \theta}{10^\circ} \right)^2
\left( \frac{k T_{\rm eff}(3 r_g)}{10 {\rm MeV}} \right)^9
\left( \frac{r_g}{10 {\rm km}} \right)^3 \quad \mbox{erg/s}.
\label{EDR2}
\end{equation}
In this restricted region, the deposited energy is free from the baryon contamination and emitted as jet.
So it will be observed as roughly factor $10^2$ magnified energy as an isotropic object. Here the efficiency of this energy to gamma ray is not clear but if we assume the efficiency as $1 \%$ \cite{Bethe}, the value of Eq.\bref{EDR2} is the deposition energy of $\gamma$ ray as an isotropic object. 
Thus, neutrino driven jet may supply the necessary power of GRB together with BZ process.
Here we have not considered electromagnetic field explicitly in these estimations. 
Let us consider electromagnetic field in $\nu\overline{\nu}$ annihilation process:

The average neutrino energy is $<\epsilon_{\nu}>\approx 10$ MeV and
the charge producing rate for Eq.\bref{EDR2} is
\be
\dot{Q}_e=e\dot{N}_{pairs}\approx 1\times 10^{37} C/s.
\ee
Whereas the magnitude of the currents in the BZ process is 
\be
I\approx \sqrt{\frac{10^{50}erg/s}{R_H}}\left(\frac{ac}{GM}\right)\left(\frac{M}{M_\odot}\right)\left(\frac{B_H}{10^{15}Gauss}\right)\approx 10^{20}~C/s,
\ee
where $R_H$ is the surface resistance of the horizon and has been set $377$ Ohm as an impedence of free space and we have set $P_{BZ}=10^{50}$ erg/s. Here we have used the conventional formula, which is sufficient for the present estimation. So it apparently seems that $\nu\overline{\nu}$ annihilation produces orders of magnitude more than enough pairs if we consider the 
electric field of Eq. \bref{EdotB} along the magnetic field.

\section{The Optical Depth Problem}
However, there remains the problem of optical depth in GRB case: That is, the source of high energy $\gamma$ may not
come out of the core of GRBs and Blazars by the reaction,
\be
2\gamma \to e^++e^-
\ee
The mean free path $l_{free}$ is estimated by
\be
l_{free}=\frac{1}{n_\gamma \sigma_{Thomp}},
\ee
where $\sigma_{Thomp}$ is the cross section of Thomson scattering,
\be
\sigma_{Thomp}=\frac{8\pi}{3}\left(\frac{e^2}{mc^2}\right)^2\approx 10^{-25}~\mbox{cm}^2.
\ee
Let us estimate $n_{\gamma}$ for GRB \cite{Murakami, Nakamura, Blandford2}.
 If we consider the radius $R$ of gas sphere of GRB
$R \approx  10^{10}$ cm and the total energy $\approx 10^{50} $ erg and photon number of $E\geq 1 MeV$
is $1 \%$, then $n_\gamma$ is roughly given by
\be
n_\gamma\approx \frac{10^{50}/10^{-6}}{(10^{10})^3}\times 10^{-2}\approx 10^{24}~\mbox{cm}^{-3}
\ee
and the mean free path $l_{free}$ is
\be
l_{free}=\frac{1}{n_\gamma\sigma_{Thomp}}\approx 
10 ~\mbox{cm} \ll R=10^{10}~\mbox{cm}
\ee
or the optical depth $\tau_{\gamma\gamma}$ satisfies
\be
\tau_{\gamma\gamma}=n_\gamma\sigma_{Thomp}R\approx \sigma_{Thomp}\frac{L\times R/c}{<\epsilon_\gamma>4\pi R^3/3}R=\frac{\sigma_{Thomp}L}{<\epsilon_\gamma>Rc}\approx 1\times 10^9,
\ee
where $L$ is the luminosity of GRB (Here we have put $c$ explicitly for clearity). Thus $\gamma$ can not escape from GRB by pair creation. In order to $\tau_{\gamma\gamma}\leq 1$, we may take the Lorentz factor $\gamma$ of the luminosity $L$ and $\sigma_{Thomp}$ into account \cite{Murakami},
\be
L:\gamma^4~~\sigma_{Thomp}:\gamma^2.
\ee
Then we know that $\gamma=O(10-10^2)$ is sufficient to solve the optical depth problem in GRBs, which may be consistent with $<\epsilon_\nu>\approx 10$ MeV adopted in the previous section.

\section{Discussion}
We have considered two scenarios on the central engines of GRBs and possibly blazars.
Two scenarios, magnetic driven (BZ) and  $\nu\overline{\nu}$ driven processes, can complement their deficit points to each other to realize consistent results. 
We have considered the homogeneous magnetic field ${\bf B}=B_0\hat{z}$ at $r\gg r_H$ under the condition of Eq.(8). We have considered the $\nu\overline{\nu}$ driven power, which needs the hot accretion disc in the equatorial plane. We are not sure how these situations are universal in general GRBs and blazars. In the present short note, we are restricted in rather qualitative arguments.  We have not discussed on the detailed spectra of jets and on their time dependence.  We will check these points quantitatively in the subsequent works.

\section*{Acknowledgments}
We would express our sincere thanks to Dr. H. Ishihara and M. Takahashi for their useful comments. This work is partly supported by JSPS KAKENHI Grant Numbers JP25H00653.

\vspace{2cm}

\end{document}